\documentclass[onecolumn,aps,12pt]{revtex4}
\usepackage{amsmath,graphicx}
\usepackage{amsmath,graphicx}
\def\,{\ifmmode\mskip\thinmuskip\else\leavevmode\thinspace\fi}

\newcommand{\dd}{\mbox{d}}
\newcommand\ba{\begin{eqnarray}}
\newcommand\ea{\end{eqnarray}}
\newcommand\nn{\nonumber}
\newcommand{\be}{\begin{equation}}
\newcommand{\ee}{\end{equation}}

\def\fun#1#2{\lower3.6pt\vbox{\baselineskip0pt\lineskip.9pt
\ialign{$\mathsurround=0pt#1\hfil##\hfil$\crcr#2\crcr\sim\crcr}}}

\begin{document}

\title{Radiative corrections to DVCS electron tensor}

\author{V.V.~Bytev}
\affiliation{Joint Institute for Nuclear Research, 141980 Dubna,
Russia}
\author{E.A.~Kuraev}
\affiliation{Joint Institute for Nuclear Research, 141980 Dubna,
Russia}

\author{E. Tomasi-Gustafsson}

\affiliation{\it DAPNIA/SPhN, CEA/Saclay, 91191 Gif-sur-Yvette
Cedex, France }

\date{\today}

\begin{abstract}
Radiative corrections in leading logarithmic approximation are calculated for 
the  differential cross section of $e^- \mu ^+$ radiative scattering. In particular the interference term due to photon emission from electron and muon blocks are calculated in helicity independent and dependent parts. The calculation is applied to the kinematical conditions of existing DVCS data in electron-proton collisions. Both helicity odd and helicity even differential cross sections are considered. 

\end{abstract}

\maketitle

\section{Introduction}
\label{sect1}

Interesting information about the Structure Functions of the proton can be found
in radiative deep inelastic electron proton scattering experiments (DIS), analyzing the
interference between the amplitudes of the radiative electron block (Bethe-Heitler amplitude)
 and the amplitudes of the radiative proton block. In the literature one can find different
suggestions for the determination of the relevant contributions to the differential cross section,
 concerning in particular lepton hadron scattering \cite{Be02}.
 
In view of the large experimental program which is underway or foreseen
at present accelerators and of the precision of the data in electron proton elastic and inelastic scattering, the necessity to achieve an adequate precision in the calculation of radiative corrections (RC) is a very actual problem. 

The theoretical description at the lowest order is based on the work of Schwinger \cite{Shwinger} and Mo and Tsai \cite{MT}. The last one contains an application to $ep$ radiative scattering to experimental data. A further improvement was given in the known paper of Yennie, Frautschi and Suura \cite{YFS}, where a simple formula was derived to describe the emission of virtual and real (soft) photons with energy lower than a value $\Delta \varepsilon$, of the order of the experimental resolution. Such photons can not be detected, in exclusive experiments. In inclusive or semi-inclusive experiments, the emission of hard, undetected photons should also be taken into account, as it escapes the detection.

The emission of an additional photon (virtual or real) is associated with a suppression factor of the order of $\alpha=1/137$, the fine structure constant. It corresponds to a small correction to the cross section, which can be estimated to $0.5\%$. However, a precise calculation of RC at higher order of perturbation theory (PT) is highly required in modern experiments at high energy. There are at least two reasons for this. Firstly, due to the emission of photons by light charged leptons, RC have an enhancement by a factor called 'large logarithm', $L=\ln Q^2/m^2$, where $Q$ is the characteristic momentum or the energy parameter and considerably exceeds the lepton rest mass $m$. Therefore the effective expansion parameter becomes $\alpha L$. Applying the general theorem about the factorization of soft and virtual photon contribution \cite{YFS}, one obtains this factor in the form:
$$W\sim b ~exp{\left [(b-1) \ln \frac{\Delta \varepsilon}{ \varepsilon}\right ]} =b\left (\frac{\Delta\varepsilon }{\varepsilon}\right )^{b-1},~b=\frac{\alpha}{\pi}(L-1),$$
where $\Delta \varepsilon $ is the energy of the photon emitted by an electron of energy $\varepsilon $.

Secondly, a kinematical effect, called 'returning' mechanism, due to hard photon emission from one of the initial charged particles may become important, in particular for processes where the cross section increases when the initial energy decreases. 

Both mechanisms were studied in lowest order of PT. Including higher orders brings, in general, large computing difficulties. However, mostly due to the study of QCD \cite{AP} processes, a powerful method was developed based on scale invariance (or renormalization group). In this frame, the behavior of the amplitudes and of the cross section can be described in the limit of vanishing lepton mass in the leading $\sim(\alpha L)^n$ (LLA) and next to leading  $\sim\alpha(\alpha L)^n$ (NLA) approximations. The application of this method to the calculation of RC provides an accuracy at thousandth level. 

The cross section including RC in LLA has the expression of the convolution of  universal functions (Structure Functions of leptons (LSF)) with a kinematically shifted cross section, calculated in Born approximation. The NLA contributions are taken into account by a $K$-factor. In this case, for two light leptons in the initial channel, one can write:
$$d\sigma(p_1,p_2,...)=\int dx_1dx_2 {\cal D}(x_1,L)  {\cal D}(x_2,L) d\sigma_B(x_1p_1,x_2p_2,...)\left (1+\frac{\alpha}{\pi}K\right ).$$
and, for the case of a single light lepton in the initial state:
$$d\sigma(p_1,...)=\int\frac{ dx {\cal D}(x,L)}{x} d\sigma_B(xp_1,...)\left (1+\frac{\alpha}{\pi}K\right ).$$
The LSF ${\cal D}(x,L)$ obeys the evolution equations of a twist 2 operator. For most quantum electrodynamic (QED) applications it is sufficient to consider only the non-singlet LSF, which has been derived in 1985 in the work of one of us \cite{KF85}.  

The motivation of the present paper is to calculate (RC) to virtual Compton
scattering (VCS) including the emission of additional hard photons as well as higher order contributions.
In particular we focus on the real part of the interference and on the total cross section.
The interference can be measured in experiments with radiative scattering of electron and positron beams on a proton target, where it is related to the difference of the corresponding cross sections in the same kinematical conditions.

Unpolarized and polarized deep virtual Compton scattering (DVCS) data are
considered to provide useful information for the extraction of the properties of generalized parton
distributions (GPD). When the accuracy of the experiment is better then 10\%, the role of radiative
corrections becomes important and a careful study of higher order contributions is mandatory.
QED radiative corrections (RC) to virtual Compton scattering on proton
($ep\to ep\gamma$) were  calculated in lowest order in Ref.
\cite{MVand}, where a detailed study of one-loop virtual corrections
including first-order soft photon emission contribution was done. Higher order RC were included by exponentiation procedure, which is valid only for small 
$\Delta\varepsilon$. 

\begin{figure}
\includegraphics[scale=1]{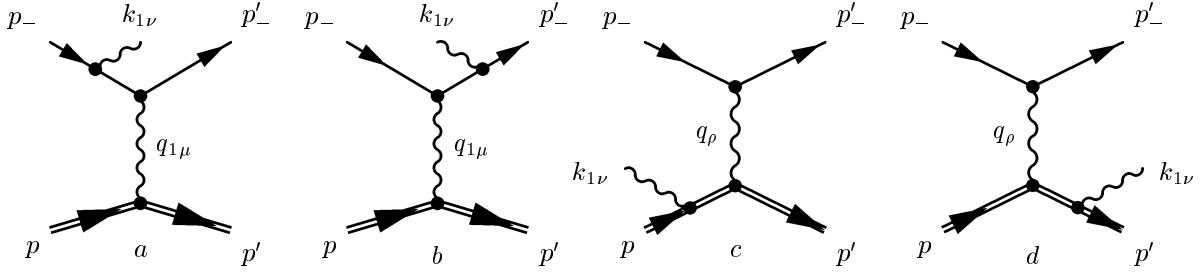}
\caption{Born Feynman diagrams for virtual Compton scattering.}
\label{fig:born}
\end{figure}

One can write schematically the cross section for the process for the DVCS process as the sum of three contributions:
\be
d\sigma^{tot}(e^- p \to e^- p\gamma) = d\sigma^{BH} + d\sigma^{DVCS}+ d\sigma^{odd},
\label{eq:eqp}
\ee
where $d\sigma ^{BH}$ is the Bethe-Heitler cross section 
(Fig. \ref{fig:born}a,b), $d\sigma^{DVCS}$
corresponds to the radiation of the photon from the proton (Fig. \ref{fig:born}c,d), and the last term corresponds to the interference between these two mechanisms. 

The paper is organized as follows.
In section II we define the kinematics and derive the formalism for the
odd part of the cross-section of the radiative $e^-\mu^+$ scattering with RC taking into account RC in the leading logarithm approximation (LLA).
In Section III we consider the contributions of three gauge invariant
classes of one-loop virtual corrections. In Sections IV and V the soft and additional hard photon emissions
in collinear kinematics are considered and the relevant generalization for all orders in LLA
in the form of electron LSF is performed.
In Section VI we extended our calculation to $ep$ scattering under realistic assumptions. We calculate the charge-even and charge-odd contributions to the  cross-section for the reactions
$e^-p\to e^-p\gamma$ and  $e^+p\to e^+p\gamma$ and the charge asymmetry, as well. The role of RC in LLA discussed.

The appendix is devoted to the kinematics of $e^-\mu^+$ radiative scattering process
in Laboratory (Lab) frame and to the parametrization of the particle four-momenta.

\section{Formalism}

Let us consider the radiative $e^-\mu^+$ scattering
\be
e^-(p_-)+\mu (p)\to e^-(p_- ')+\mu (p^{'})+\gamma(k_1)
\label{eq:eq1}
\ee
as a model for DVCS in electron proton radiative scattering, considering the muon as a structureless proton.
The contribution to the differential cross section of reaction (\ref{eq:eq1}), which corresponds to the so called up-down interference of the amplitudes describing the radiation from the electron
and the muon blocks, in the lowest order of PT, can be written as
\begin{gather}
(\dd\sigma)^{e\mu\gamma}_{odd}=\frac{4(4\pi\alpha)^3}{stt_1}H_{\mu\nu\rho}E_0^{\mu\nu\rho} \dd\Gamma,
\nonumber \\
\dd\Gamma=\frac{\dd^3p_- ' }
{2\varepsilon_- '} \frac{\dd^3p '}{2\varepsilon '}
\frac{\dd^3k}{2\omega}\frac{\delta^4(p_-+p-p_- '-p '-k_1)}{(2\pi)^5}.
\end{gather}
$p_- '$ and $\varepsilon_- '$ ( $p_-$ and $\varepsilon_-$) are the momentum and the energy of the scattered electron (muon). The odd DVCS tensors for electron and muon are:
\begin{gather}
E_0^{\mu\nu\rho}(p_-,k_1,p_- ')=\frac{1}{4}
Tr\,\, \hat{p}_-^{'}\left ({\gamma}^\nu\frac{\hat{p}_-'+\hat{k_1}}{\chi_-'}{\gamma}^\mu
-\gamma^\mu\frac{\hat{p}_--\hat{k}_1}{\chi_-}
\gamma^\nu\right )\hat{p}_-\gamma^\rho \, ,
\nn \\
H_{\mu\nu\rho}=\frac{1}{4}
Tr~
(\hat{p}'+M)\left ({\gamma}_\rho\frac{\hat{p}-\hat{k_1}+M}{ -\chi }{\gamma}_\nu
+{\gamma}_\nu\frac{\hat{p'}+\hat{k_1}+M}{\chi ' }{\gamma}_\rho \right )(\hat{p}+M){\gamma}_\mu .
\end{gather}
The on-mass shell conditions and kinematics invariants are defined:
\begin{gather}
 p_-^2= p_-^{'2}=m^2, \quad k_1^2=0,\quad p^2=p'^2=M^2,
\nn  \\
\chi_-=2k_1 p_-, \quad \chi_-'=2k_1 p_-',\quad
{\chi}=2k_1 p, \quad {\chi}'=2k_1 p',
\nn  \\
s=2p_-p,\quad s_1=2p_-'p', \quad t=-Q^2=-2p_- p_-',
 \nonumber \\
 t_1=q_1^2=2M^2-2p p',\quad u=-2p_-p',\quad u_1=-2p_-'p,
\nonumber \\
s+s_1+t+t_1+u+u_1=0,
\end{gather}
where $m$ and $M$ are electron and  muon (proton) mass.
Throughout the paper we will suppose
\be
\quad s\sim s_1\sim -t\sim -t_1\sim-u\sim-u_1\sim\chi_-\sim\chi_-^{'}\sim\chi\sim\chi^{'}\gg m^2,
\ee
and we will systematically omit terms of the order of $m^2/s$ compared to those of order of unity. This kinematical region corresponds to large-angle final particle emission in Lab frame, where the calculation is performed.

In order to make the comparison with the experimental data, we chose the following set of four independent variables:
\begin{gather}
Q^2=-t, \quad t_1,\quad x_{Bj}=\frac{Q^2}{2pq},\quad q=p_--p_-'~\mbox{~and~}\phi,
\end{gather}
where $\phi$ is the azimuthal angle between the plane containing the three-momenta of the initial and the scattered electrons
$(\vec{p}_-,\vec {p}_-^{\,\,\,'})$ and the hadronic plane, containing the momentum transfer to the electron, $\vec{q}$, and
the scattered muon momentum $\vec{p}^{\,\,\,'}$ \cite{Ba04}.

The phase volume can be rewritten in terms of these variables (see details in the Appendix)
as:
\begin{gather}
\dd\Gamma=\frac{\dd\Phi_4}{2^8\pi^4R}, \quad
\dd\Phi_4=\frac{1}{sx_{Bj}}\dd\phi\dd Q^2\dd t_1\dd x_{Bj}, \quad
R=\biggl[1+\frac{4M^2x_{Bj}^2}{Q^2}\biggr]^{\frac{1}{2}}.
\label{eq:volume}
\end{gather}
The Born cross section (in the lowest order of perturbation theory  (PT)) has the form
\be
(\dd\sigma)^{e\mu\gamma}_{odd}=\frac{\alpha^3}{2\pi stt_1R}W\dd\Phi_4,\quad
W=2H_{\mu\nu\rho}E^{\mu\nu\rho}_0.
\ee
In case of massless muon we recover the result from Ref. \cite{Berends}:
\be
W_{M=0}= (s^2+s_1^2+u^2+u_1^2)
\biggl[
\frac{s}{\chi_- \chi }+
\frac{s_1}{ \chi_- '\chi '}
+\frac{u}{\chi ' \chi_-}+
\frac{u_1}{\chi_- '\chi } \biggr].
\ee
Below we consider the radiative corrections to this part of differential cross-section.
We show that when the energy fraction of the scattered electron
is not fixed, we
obtain in LLA ( 
$\alpha/\pi\ll 1$, $\alpha/\pi L\sim1$, and $L=\ln(Q^2/m^2)$ is the large logarithm):
\ba
\frac{\dd\sigma^{e\mu\gamma}_{odd}}{\dd\Phi_4}&=&\frac {\alpha^3}{2\pi sQ^2 t_1}
\int_{x_0(\phi)}^1  \displaystyle\frac {\dd x} {x}{\cal D}(x,L)\displaystyle\frac{W(x)}{[1-\Pi(xt)][1-\Pi(t_1)]}
\Psi(x),\nn \\
~\Psi(x)&=& \displaystyle\frac{1}{R'I}\biggl[1-\displaystyle\frac{sx_{Bj}(1-x)}{Q^2}\biggr ]^{-1},
\label{eq:cross:D}
\ea
where $\Pi(Q^2) $ is the contribution to vacuum polarization from the
light lepton (electron), $W(x)=W(p_-\to p_-x)$ and $R'$, $I$ are defined in the Appendix.
$D(x,L)$ is the non-singlet LSF of the electron \cite{KF85}
\begin{gather}
 {\cal D}(x,L)= \frac{1}{2}\beta(1-x)^{\beta/2-1} \left [ 1+\frac{3}{8}\beta\right ]
-\frac{1}{4}\beta(1+x) + {\cal O}(\beta^2),\quad \beta= 2\frac{\alpha}{\pi} (L-1).
\end{gather}
The physical requirements $\varepsilon_-'>0$ and the on-mass shell condition for the real photon lead to the restrictions:
\be
x> x_0(\phi), \quad 1-\frac{Q^2}{sx_{Bj}}>0.
\label{eq:eq10}
\ee
The determination of the quantity $x_0(\phi)$ is given in the Appendix.
The helicity dependent part of DVCS cross section on proton is
\be
\frac{d^4\Sigma}{d\phi}=\frac{1}{2}\left (\frac{d\sigma^{\rightarrow}}{d\phi}-  \frac{d\sigma^{\leftarrow}}{d\phi}\right )
\label{eq:asym}
\ee
and it is sensitive to the imaginary part of the DVCS amplitude. 
Let us calculate the proton Compton amplitude in the structureless approximation, and parametrize the nucleon structure by a general factor $G$.

The relevant part of the matrix element squared can be written as:
\be
\Delta|M^{\rightarrow}|^2-\Delta|M^{\leftarrow}|^2\sim Im(G)~[\vec p_- \times \vec p~'_-]\cdot \vec k ~{\cal F},
\label{eq:dm}
\ee
with 
\ba
{\cal F}&=&(2t+4m^2)\left (\frac{1}{\chi_1\chi_2}+\frac{1}{\chi_2\chi_1'}\right ) 
+2(s-u_1)\left(\frac{1}{\chi_1\chi_2}-\frac{1}{\chi_2\chi_1'}\right ) \nn\\
&&
- 2\chi_2\left (\frac{1}{\chi_1\chi_2'}+\frac{1}{\chi_1'\chi_2'}\right ) 
+\frac{4(s-M^2)}{\chi_1\chi_2}+\frac{4(u_1-M^2)}{\chi_1'\chi_2'}.
\label{eq:fm}
\ea

\section{One-loop virtual corrections}
\begin{figure}
\includegraphics[scale=1]{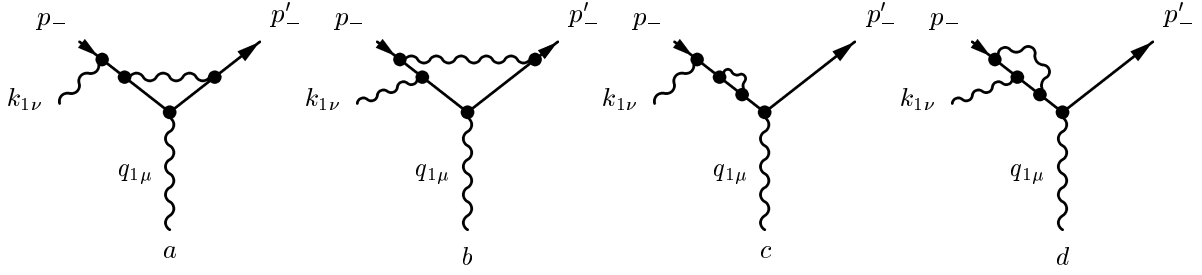}
\caption{Some one-loop FD for virtual Compton scattering.}
\label{fig:corr}
\end{figure}

In LLA only Feynman diagrams (FD) where a single photon is transferred between the
muon and the  electron blocks contribute to cross-section (see Fig. \ref{fig:corr}).
In our considerations we omit FD with two virtual exchanged photons due to
the cancellation of such  contributions when one includes the amplitude corresponding to soft
photon emission between electron and muon blocks. The details of this 'up-down cancellation', which holds in LLA, were discussed in \cite{Ku06} and Refs. therein. The corresponding contribution goes beyond the limits of accuracy of the present calculation.

\begin{figure}
\includegraphics[scale=1]{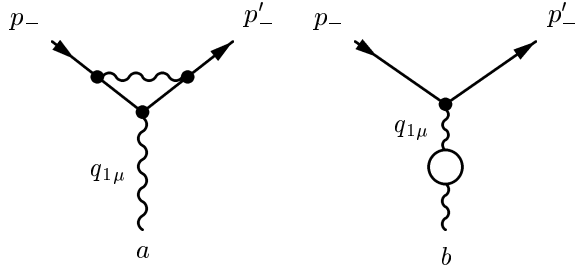}
\caption{Dirac and vacuum polarization contribution for one-loop FD.}
\label{fig:Dirac}
\end{figure}

In the calculation, only FD drawn in Fig. \ref{fig:corr} can be considered. The corresponding part of the total matrix element is denoted as $M^{\gamma}$.
The total contribution to the DVCS tensor can be restored from
the interference of these amplitudes (Fig. \ref{fig:corr}) with the Born one (Fig. \ref{fig:born}c or d):
\be
E_{\mu\nu\rho}^{virt}=[1-P(p_-\leftrightarrow -p_-^{'})]M_{\mu\nu}^\gamma(M_\rho)^\star .
\label{eq:eq12}
\ee
The matrix element describing the electron self-energy (see Fig. \ref{fig:corr}c,d)
and the vertex function of the real photon emission by the initial electron have the form \cite{KMF87}:
\begin{gather}
\frac{\alpha}{2\pi}\overline{u}(p_-^{'})\gamma_\mu\biggl[A_1\biggl(\hat{e}-\hat{k_1}\frac{ep_-}{k_1p_-}\biggr)
+A_2\hat{k_1}\hat{e}\biggr]u(p_-).
\end{gather}
The contribution of the structure $A_1$ disappears in the limit $m\to 0$ \cite{tables},
whereas $A_2$ survives, providing the following contribution to the DVCS tensor:
\begin{gather}
E_{\mu\nu\rho}^{virt_1}=\frac{\alpha}{\pi}\frac{1}{\chi_-}\left (\ln\frac{\chi_-}{m^2}-\frac{1}{2}\right )
Tr\,\, \hat{p}_-^{'}{\gamma}_\mu\hat{k_1}{\gamma}_\nu\hat{p}_-{\gamma}_\rho\, ,
\end{gather}
The contributions of the virtual photon emission vertex of type FD (Fig. \ref{fig:corr}a) as well as of the box-type (Fig. \ref{fig:corr}b)
have the form:
\begin{gather}
E_{\mu\nu\rho}^{virt_2}=\frac{\alpha}{4\pi}\int\frac{\dd^4k}{i\pi^2}
\biggl\{\frac{S_1}{-\chi_-}+\frac{S_2}{(p_--k)^2-m^2}\biggr\}
\frac{1}{(k^2-\lambda^2)[(p_-'-k)^2-m^2][(p_- -k_1-k)^2-m^2]},
\end{gather}
where
\begin{gather}
S_1=
\frac{1}{4}Tr\,\, \hat{p}_-^{'}{\gamma}_\lambda(\hat{p}_-^{'}-\hat{k}){\gamma}_\mu
(\hat{p}_-^{'}-\hat{k}_1-\hat{k}){\gamma}_\lambda(\hat{p}_--\hat{k}_1){\gamma}_\nu
\hat{p}_-{\gamma}_\rho\, ,
\\ \nonumber
S_2=
\frac{1}{4}
Tr\,\, \hat{p}_-^{'}{\gamma}_\lambda(\hat{p}_-^{'}-\hat{k}){\gamma}_\mu
(\hat{p}_-^{'}-\hat{k}_1-\hat{k}){\gamma}_\nu(\hat{p}_--\hat{k}){\gamma}_\lambda
\hat{p}_-{\gamma}_\rho\, .
\end{gather}
Their calculation requires scalar, vector and tensor  (up to rank three)
integrals with three and four denominators, which are listed in \cite{tables}.

Both $E_{\mu\nu\rho}^{virt_1}$ and $E_{\mu\nu\rho}^{virt_2}$ do not satisfy gauge invariance.
Only the right hand side of the expression (\ref{eq:eq12}) restore the property of gauge invariance.

After applying (\ref{eq:eq12}), the sum of the vertex contributions excluding FD
with Dirac form-factor (see Fig. \ref{fig:Dirac}) are:
\begin{gather}
E_{\mu\nu\rho}^{virt}=E_{\mu\nu\rho}^{virt_1}+E_{\mu\nu\rho}^{virt_2}
=E_{\mu\nu\rho}^{0}\frac{\alpha}{\pi}\biggl[-\frac{1}{4}L^2
+\frac{1}{2}\ln\frac{m ^2}{\lambda^2}(1-L)+\frac{3}{4}L\biggr],\quad L=\ln\frac{-q^2}{m^2}.
\end{gather}
In this expression it was assumed that all terms proportional to
$k_{1\nu}$ give a vanishing contribution, due to the Lorentz condition  $e(k_1)k_1=0$.

\section{Soft photon emission and Dirac form-factor contributions}
Finally let us consider the vertex-type corrections to the electron scattering vertex
without real photon emission (see Fig. \ref{fig:Dirac}a) and the contribution of additional
soft photon emission with energy not exceeding  $\Delta\varepsilon$.

Both contributions are proportional to the Born DVCS terms:
\begin{gather}
E_{\mu\nu\rho}^{soft+D}=E_{\mu\nu\rho}^{0}\biggl(\frac{\alpha}{\pi}\Gamma_1(q^2)+\delta_{soft}\biggr),
\\ \nonumber
\delta_{soft}=-\frac{4\pi\alpha}{(2\pi)^3}\int\frac{\dd^3k_2}{2\omega_2}
\biggl(\frac{p_-}{p_-k_2}-\frac{p_-^{'}}{p_-^{'}k_2}\biggr)^2\biggl|_{\omega_2\ll\Delta\varepsilon},
\end{gather}
where
\begin{gather}
\frac{\alpha}{\pi}\Gamma_1(q^2)=\frac{\alpha}{\pi}
\biggl[\ln\frac{m}{\lambda}(1-L)-\frac{1}{4}L^2
+\frac{3}{4}L+\frac{\pi^2}{12}-1\biggr],
\\ \nonumber
\delta_{soft}=\frac{\alpha}{\pi}
\biggl[(L-1)\ln\frac{(\Delta\varepsilon)^2m ^2}{\lambda^2\varepsilon_-\varepsilon_-^{'}}
+\frac{1}{2}L^2-\frac{1}{2}\ln^2\frac{\varepsilon_-^{'}}{\varepsilon_-}
-\frac{\pi^2}{3}+\mathrm{Li}_2\left (\cos^2\frac{\theta}{2}\right )\biggr],
\end{gather}
where $\varepsilon_-$ is the energy of the incident electron and $\theta$ is electron scattering angle.

Combining all contributions containing large logarithms,
we arrive to the lowest order expansion of the right hand side,  which does not contain the auxiliary
parameter $\lambda$.  Omitting the terms of order of unity we
obtain:
\begin{gather}
E_{\mu\nu\rho}^{summed}=E_{\mu\nu\rho}^{virt}+E_{\mu\nu\rho}^{soft+D}=E_{\mu\nu\rho}^{0}
\frac{\alpha}{\pi}\biggl[\ln\frac{(\Delta\varepsilon)^2}{\varepsilon_-\varepsilon_-^{'}}
+\frac{3}{2}\biggr](L-1).
\end{gather}

\section{Additional hard photon emission contribution}

The contributions arising from the emission of an additional hard photon with energy $\omega_2>\Delta\varepsilon$ can be written in form of two terms. The first one, corresponding to collinear kinematics,
contains a large logarithm of type $L$ and
can be calculated with the help of the quasi real electron method \cite{BFK}.
It has a form:
\begin{gather}
\frac{\alpha}{2\pi}\int\limits_{x_0(\phi)}^{1-\Delta_1}\dd x [P(x)(L_1 -1)+1-x] E_{\mu\nu\rho}^{0}(p_-x,p_-^{'},k_1),
\end{gather}
for the case of photon emission close to the initial electron, and
\begin{gather}
\frac{\alpha}{2\pi}\int\limits_{y(1+\Delta_2)}^{1}\frac{\dd z}{z}
\left [P\left (\frac{y}{z}\right )( L_2-1)+1-\frac{y}{z}\right ]
E_{\mu\nu\rho}^{0}\left(p_-,\frac{z}{y}p_-',k_1\right ),
\end{gather}
 for the case of photon emission close to the scattered electron with
\begin{gather}
\Delta_1=\frac{\Delta\varepsilon}{\varepsilon_-} , \quad
\Delta_2=\frac{\Delta\varepsilon}{\varepsilon_-^{'}}, \quad
P(z)=\frac{1+z^2}{1-z},
\end{gather}
with
\be
L_1=\ln\frac{ \varepsilon_-^2\theta_0^2}{m^2},~
L_2=\ln\frac{ \varepsilon_- '^2\theta_0^2}{m^2},~
\ee
This contribution arises when the photons are emitted in a narrow cone, within an angle $\theta_0\ll 1$, along the directions of the initial and the scattered electrons.

The contribution from non collinear kinematics $\theta > \theta_0$ cancels the $\theta_0$ dependence and does not contain large logarithms. Omitting non leading terms, we can write $L_1=L_2=L$.

By summing up all contributions, we can put the
cross section of the radiative production in the form:
\ba \label{18}
&& E_{\mu\nu\rho}(p_-,p_-^{'},k_1)
= \int\limits_{0}^{1}\dd x D(x,L)\int\limits_{y}^{1} \frac{\dd z}
{z}D(\frac{y}{z},L)
\frac{E_{\mu\nu\rho}^{0}(xp_-,\frac{z}{y}p_-^{'},k_1)}
{q^2(x,z)q_1^2}
\frac{1}{[1-\Pi(q^2(x,z)][1-\Pi(q_1^2)]}
\nn \\  &&
D(x,L)=\delta(1-x)+\frac{\alpha}{2\pi}P^{(1)}(x)(L-1)+....,
\;
\\ \nonumber
&&P^{(1)}(x)=\lim_{\Delta\to 0}\biggl[(2\ln\Delta+\frac{3}{2})\delta(1-x)
+\Theta(1-x-\Delta)\frac{1+x^2}{1-x}\biggr]
=\biggl(\frac{1+x^2}{1-x}\biggr)_+ \,\, .
\ea
Here $1/ [1-\Pi(q^2(x,z)]$ is the polarization vacuum factor (see Fig. \ref{fig:Dirac}b),
  $\Pi(t)\sim \frac{\alpha}{3\pi} (L-\frac{5}{3})$ and $ q^2(x,z)=q^2xz/y$.

This expression is in agreement with the result previously obtained for the whole
differential cross-section in Ref. \cite{eemumugamma}
where the RC to the muon block were also taken into account.
Performing the integration on the scattered electron energy fraction $y$ and using
the normalization property of the LSF
\begin{gather}
\int\limits_0^1\dd zD(z,L)=1
\end{gather}
we recover the expression (\ref{eq:cross:D}).

The differential cross section for the reaction (\ref{eq:eq1}) in LLA can therefore be
expressed in terms of the shifted Born cross section as \cite{eemumugamma}:
\be
\dd\sigma^{e^\pm\mu\to e^\pm\mu\gamma}(p_{\pm},...)=\int  \displaystyle\frac{{\dd x} {\cal D}(x,L)}{[1-\Pi(xt)][1-\Pi(t_1)]} d\sigma_B^{e^\pm\mu\to e^\pm\mu\gamma}(xp_{\pm},..),
\label{eq:eqa}
\ee
with the following expression :
\ba
\dd\sigma_B^{e^\pm\mu\to e^\pm\mu\gamma}(p_\pm,....)&=& \displaystyle\frac{2^7\pi^3\alpha^3}{stt_1}
\left ( s^2+s_1^2+u^2+u_1^2\right )\nn \\
&\times&
\left [ -\displaystyle\frac{t_1}{\chi_-\chi_-'} -\displaystyle\frac{t}{\chi\chi '}
\mp \left (\displaystyle\frac{u}{\chi_-\chi '}+\displaystyle\frac{u_1}{\chi_-'\chi }+\displaystyle\frac{s}{\chi_-\chi }
+\displaystyle\frac{s_1}{\chi_-'\chi '} \right )\right ]
\dd\Gamma 
\label{eq:eqb}
\ea
for the non-shifted cross-section. The explicit expression for the shifted cross section is derived in a straightforward way, by replacement of the shifted kinematics.

\section{Numerical Calculation. Application to $ep$ DVCS}

Let us consider the case of unpolarized electron and unpolarized proton target and give an
estimation of the RC to the cross section calculated in the Born approximation. We consider,
in particular the calculation for the reaction (\ref{eq:eq1}) as a model for $ e^\pm+ p\to e^\pm +p+ \gamma $,
replacing the muon mass by the proton one.

\begin{figure}
\includegraphics[scale=.6]{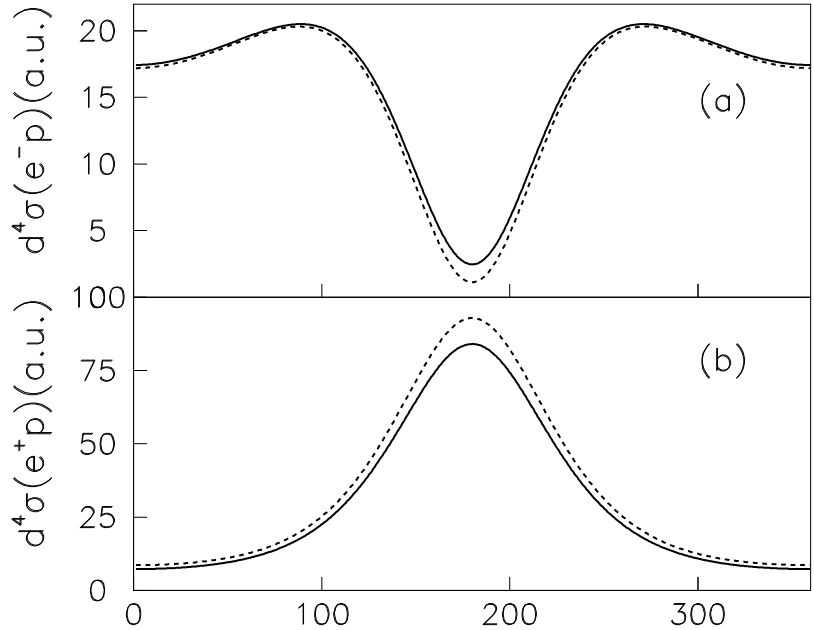}
\caption{(a):Calculation of the differential cross section for  $e^- p\to e^-p \gamma$ (i.e., $e^-\mu\to e^-\mu\gamma$ with $M_{\mu}=$1 GeV) for the kinematics corresponding to \protect\cite{Mu06}: $Q^2=2.3 ~$ GeV$^2$, $-t_1$=0.28 GeV$^2$, $x_{Bj}$=0.36 (solid line). The result after applying radiative correction is also shown (dashed line); (b): same for positron scattering.}
\label{fig:sigth}
\end{figure}

The four-fold differential cross section, $d^4\sigma(\phi)$ has been calculated according to Eqs. (\ref{eq:eqa},\ref{eq:eqb}) for kinematical conditions as in Ref. \cite{Mu06}. The results for electron (a) and positron (b) scattering are shown in Fig. \ref{fig:sigth} before (solid line) and after (dashed line) applying radiative corrections. One can see that at $\phi=\pi$ the cross section for electrons (positrons) has a minimum (maximum) and that RC induce a $\phi$ dependent relative correction. 

The calculated relative effect may be applied to the experimental data. In Ref. \cite{Mu06} RC were calculated for $e^-+p\to e^-+p+\gamma$ following Ref. \cite{MVand} and applied to the data with the help of a Monte Carlo simulation. This procedure
resulted in a correction of the yield by a factor $F=0.91\pm 0.02$ which is constant with respect to $\phi$, convoluted with $\Delta \varepsilon$ dependent corrections, which were included in a Monte Carlo simulation together with acceptance corrections. The overall effect was to increase the experimental yield of about 20\%, roughly constant with $\phi$. 

In case of $e^-p$, LLA radiative corrections induce on one side a lowering of the cross section, with respect to the calculated Born cross section and on the other side, a change of the $\phi$ dependence. This strong $\phi$ dependence is an effect of hard photon emission. In an exclusive measurements, where the four momenta of all the particles involved are precisely determined, the importance of this effect could be quantitatively determined.

\begin{figure}
\includegraphics[scale=.6]{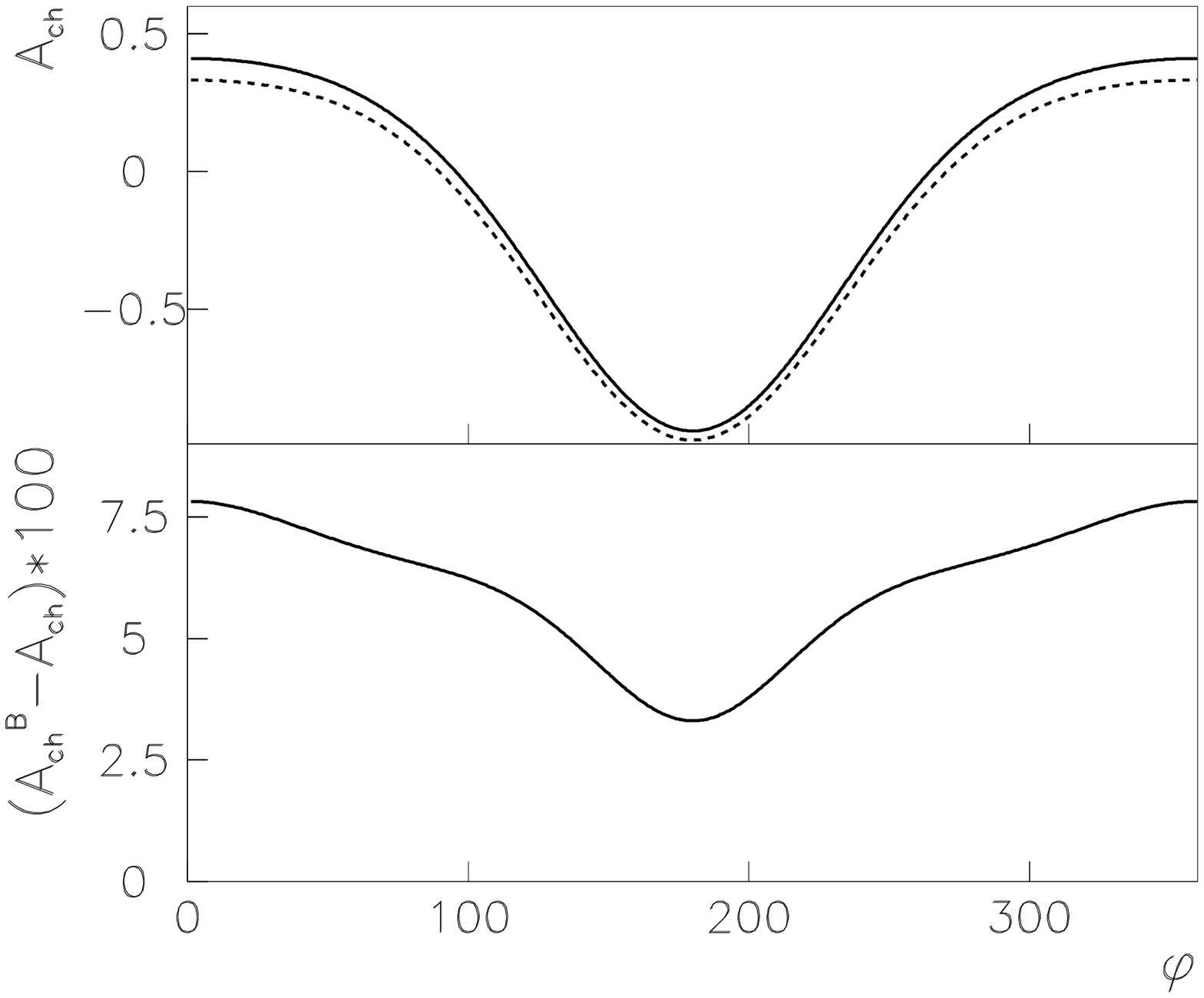}
\caption{ Calculation of the charge asymmetry (Eq. \protect\ref{eq:eqb}) in Born (solid line) and
 LLA  (dashed line) approximation (top). The relative value in percent is also drawn (bottom). Same kinematics as in Fig.  \protect\ref{fig:sigth}.}
\label{fig:asym}
\end{figure}

Let now consider the charge asymmetry:
\be
A_{ch}=\displaystyle\frac{d\sigma^{e^-\mu\to e^-\mu\gamma} -d\sigma^{e^+\mu\to e^+\mu\gamma}}
{d\sigma^{e^-\mu\to e^-\mu\gamma} +d\sigma^{e^+\mu\to e^+\mu\gamma}}.
\label{eq:as}
\ee
We can consider the calculation of $A_{ch}$  as a model for radiative $ep$ scattering
(after replacing the muon mass with the proton mass). In Born and LLA approximation  $A_{ch}$ is shown in Fig. \ref{fig:asym} (top),
 and the relative difference in Fig. \ref{fig:asym} (bottom). 

The charge asymmetry is large, and may exceed 0.5 for in plane kinematics. Radiative corrections are of the order of 5\% with a smooth $\phi$ dependence. 
This quantity is especially interesting as it is in principle measurable at electron positron rings with fixed target.

\begin{figure}
\includegraphics[scale=.6]{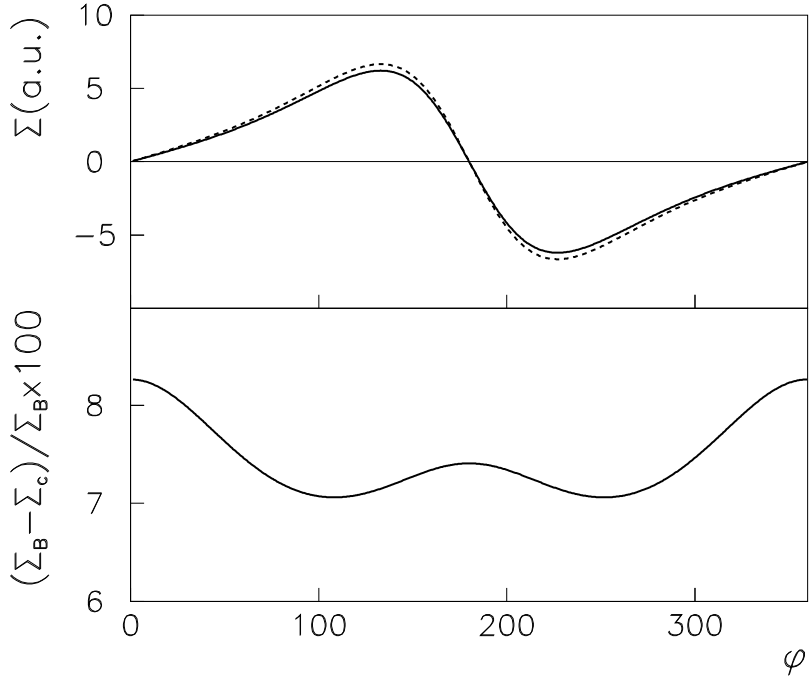}
\caption{ Helicity asymmetry in arbitrary units (top):Born calculation (solid line), radiatively corrected (dashed line). Relative value of the corrections in percent (bottom). Same kinematics as in Fig.  \protect\ref{fig:sigth}.}
\label{fig:heli}
\end{figure}

Let us calculate the helicity dependent cross section Eq. (\ref{eq:asym}) and the radiative corrections, calculated in LLA as a function of $\phi$. The result is shown in Fig. \ref{fig:heli}. As expected, we obtain an antisymmetric function, which can expanded in harmonics by $\sin\phi,~\sin2\phi ...$, which coefficients have physical meaning of all order twist contributions.

The radiative corrections to the helicity dependent cross section are of the order of several percent, with a small modulation in $\phi$.

\section{Conclusions}

We calculated radiative corrections to VCS in the high-energy limit. The emission of hard photon in collinear kinematics is also included.
The sum of all contributions (including soft photon emission) does not depend neither on the fictitious photon mass $\lambda$ or on the soft photon energy $\Delta \varepsilon$, and it is consistent with the renormalization group prediction.

We applied the calculation, which is rigorous for the $\mu$ case, to proton scattering, after correcting for the mass. The proton structure can be taken into account in terms of electromagnetic form factors, which are function of $t_1$ and are not influenced by the conversion procedure to the shifted kinematics. However, let us note that taking into account nucleon form factors may violate the current conservation condition \cite{Kr65}. A self-consistent procedure requires a {\it ad hoc} modification of the nucleon propagator. This can be done including the excited states of the nucleon, such as the $\Delta$ resonance \cite{Ko05,By07}. It appears that elastic and inelastic processes partly compensate the effects of the strong interaction. Based on arguments of analyticity and unitarity \cite{BFK}, one can expect that, taking into account the complete set of inelastic states in the intermediate state of the virtual Compton amplitude, an almost complete cancellation takes place, up to the contribution of structureless proton. This is the reason for which the approximation of structureless proton can be considered realistic. Moreover, if one builds the relevant ratios, such as $A_{ch}$, the effect of form factors is essentially canceled.

The effect of hard photon emission is considerable, and the 'returning mechanism' which is essentially expressed in
form of convolution of the shifted Born cross section with the electron LSF, may become important. At our knowledge, such mechanism was not considered in the previous literature for the reaction under consideration here.

Comparing with the scheme adopted to correct the experimental data, (i.e. taking into account first order RC, partly calculated with the help of a Monte Carlo and partly applying a constant factor to the final results) the present approach suggests a $\phi$ dependent correction, mostly due to hard photon emission. The importance of this effect could be tested in a truly exclusive experiment and it may affect the extraction of the physical information from the Fourier analysis of the $\phi$ dependence of the relevant observables.

\section{Acknowledgments}
Two of us (V.V.B and E.A.K.) are grateful to Saclay Institute of Physics for hospitality.
We also grateful to INTAS grant 05-1000-008-8323 for financial support and grant MK--2952.2006.2. Thanks are due to C. Munoz-Camacho for providing us with tabulated values and details on the experiment, and to D. Marchand, D. L'Huillier, P. Guichon and J. Van de Wiele for useful discussions.

\section{Appendix}
The emission of collinear photons from the initial electron induces a shift of the kinematical invariants and of the phase volume.

Firstly, let us consider the kinematics without photon emission (non shifted).
Introducing the set of  new variables defined above and after performing the integration
on the photon variables one has:
\begin{gather}
\dd\Gamma=\frac{1}{(2\pi)^5}\frac{\varepsilon_-'\dd\varepsilon_-'p'^2\dd p'}{4\varepsilon'}
\dd O_- \dd O' \delta((q+q_1)^2),
\end{gather}
Using the the Lab frame ($\vec{p}=0$, $\vec{q}_1=-\vec{p}^{\,\,\,'}$) and choosing
the $z$ axes along the direction of $\vec{q}$,  we obtain:
\begin{gather}
\dd O_- \dd O_-'\delta((q+q_1')^2)=2\pi\dd c \dd c' \dd\phi \delta(t_1-Q^2+2q_0q_{10}-2|\vec{q}|p' c')
=\frac{\pi}{|\vec{q}|p'}\dd c \dd\phi,
\end{gather}
with  $c=\cos\theta$, $\theta$ is the angle between incident and outgoing electron momenta $\vec{p}_-$ and
$\vec{p}_-^{\,\,\,'}$, and $c'=\cos\theta'$ is the cosine of the angle between $\vec{q}$ and $\vec{p}^{\,\,\,'}$.
Using the definitions
\begin{gather}
Q^2=2\varepsilon_-\varepsilon_-'(1-c),
\quad
t_1=M^2-2M\varepsilon',
\end{gather}
we can obtain
\begin{gather}
\dd\varepsilon_-'=\frac{Q^2}{2M}\frac{\dd x_{Bj}}{x_{Bj}^2},
\quad
\vec{q}^{\,\,2}=q_0^2+Q^2,
\quad
q_0=\varepsilon_--\varepsilon_-'=\frac{Q^2}{2Mx_{Bj}},
\nn \\
\dd\varepsilon'=\frac{\dd t_1}{2M},
\quad
\dd c=\frac{\dd Q^2}{2\varepsilon_-\varepsilon_-'},
\quad
|\vec{q}|=\frac{Q^2 R}{2x_{Bj}M}.
\end{gather}
and $R$ is given in (\ref{eq:volume}),
After some algebra we obtain the phase volume in terms of the new variables:
\begin{gather}
\dd\Gamma = \frac{(2\pi)^{-4}}{16R} \dd\Phi_4,
\quad
\dd\Phi_4=
\frac{\dd x_{Bj} \dd t_1 \dd Q^2 \dd\phi}{s x_{Bj} },~s=2M\varepsilon_-.
\end{gather}
Let us chose the $y$ axis transverse to the electron scattering plane $\vec{n}_y\parallel\vec{p}_-^{\,\,\,'}\times\vec{p}_-$
and the $x$ axis $\vec{n}_x\parallel\vec{q}\times\vec{n}_y$. So we can parameterize the 4-vectors as $a=(a_0,a_z,a_x,a_y)$ :
\begin{gather}
p_-=\varepsilon_- \{1,c_-,s_-,0\}, \quad
p_-'=\varepsilon_-'\{1,c_-',s_-', 0\},
\nn  \\
q=\{q_0,|\vec{q}|,0,0\},
\quad
p'=\{\varepsilon',p'c',p's'\cos\phi, p's'\sin\phi\}.
\end{gather}
From the conservation law $\vec{p}_-=\vec{p}_-^{\,\,\,'}+\vec{q}$, $\vec{q}=\vec{k}_1+\vec{p}^{\,\,\,'}$ we obtain:
\ba
c'&=&\frac{Mx_{Bj}}{{p}^{\,\,\,'}R}\left (1-\frac{t_1}{Q^2}-\frac{t_1}{2M^2x_{Bj}}\right ),\quad
s'=\sqrt{1-c'^2},\nn \\
c_-&=&\frac{1}{R}\left (1+\frac{2M^2x_{Bj}}{s}\right ),\quad
s_-=\sqrt{1-c_-^2},\nn \\
c_-'&=&\frac{1}{R}\left (1-\frac{2M^2x_{Bj}^2}{sx_{Bj}-Q^2}\right ),\quad
s'_-=\sqrt{1-c_-^{'2}}.
\ea
The energy of the scattered electron is:
\be
\varepsilon_-'=\varepsilon_- \left (1-\displaystyle\frac{Q^2}{sx_{Bj}}\right ) >0.
\label{eq:limits}
\ee
So the variables $Q^2$, $s$, $x_{Bj}$ must obey the condition (\ref{eq:limits}).
All the kinematical invariants can be expressed in terms of the variables:
$u$, $u_1$ and $s_1$ as follows:
\be
\chi_-=s-Q^2+u,~\chi_-'=Q^2-u_1-s_1,~\chi '=-t_1-u-s_1,~
\chi =2M^2+\frac{Q^2}{x_{Bj}}- 2M \varepsilon ',
\label{eq:eqs1}
\ee
with
\ba
s_1&=&2p_-'p'=2\varepsilon_- '[ \varepsilon '-p'(c_-'c'+ s_-'s'\cos\phi)],\nn \\
u_1&=&-2pp_-'=-2\varepsilon_- ' M,\quad
u=-2p_-p'=-2\varepsilon_- [ \varepsilon '-p'(c_-c'+ s_-s'\cos\phi)].
\label{eq:eqs2}
\ea
The remaining variables are
\be
\varepsilon '=M-\frac{t_1}{2M}, \quad p'=\sqrt{\varepsilon '^2-M^2}.
\ee
Let us consider now the shifted kinematics, which consists in the replacement
$q\to q_x=p_-x-p_-'$. It is convenient to introduce a shifted Bjorken variable:
$x_{Bj}'= \frac{xQ^2}{2Mq_{x0}}$, with the following relation:
$$\frac{x}{x_{Bj}'}=\frac{1}{x_{Bj}}-\frac{s(1-x)}{Q^2}.$$
Particular attention should be devoted to the calculation of the integral on the variable $c'$:
\be
Y = \int dc'\delta[-xQ^2+t_1+2q_{x0}q_{10}+2\vec q_x \vec{p}^{\,\,\,'}],
\label{eq:delta}
\ee
which arises due to the fact that the direction of $\vec{q}_x$ does not coincide with the direction of $\vec q$.
The result of the integral (\ref{eq:delta}) can be written as:
\be
Y =\frac{1}{2|\vec{q}_x|p'}\frac{1}{I},\quad I=\left |\frac{dc'_x}{dc'}\right |,\quad
c'_x=c'(x)c_x+s'(x)s_x\cos\phi,\quad  c_x=\cos\widehat{\vec q\,\,\, \vec q_x},
\label{eq:eqcpr}
\ee
where $c_x$ can be written as:
$$
c_x=\frac{1}{RR'}\left [1+\frac{2M^2x_{Bj}x'_{Bj}(1+x)}{Q^2x} \right ],\quad
R'=\sqrt{1+\frac{4M^2 x'^2_{Bj}}{xQ^2}},\quad s_x=\sqrt{1-c_x^2}.
$$
In Eq. (\ref{eq:eqcpr}) we introduce the notation $c'(x)$, $s'(x)$ for the
correspondent variables of $c'$ and $s'$ in shifted kinematics.

Keeping in mind the two possible solutions of Eq. (\ref{eq:eqcpr}) $c_{\pm}'(x)$, $s_{\pm}'(x)$, 
the quantity $I$ must be understood as
\be
\frac{1}{I}=\frac{|c'_xs_x\cos\phi-c_x\sqrt{{\cal D}_0}|+|c'_xs_x\cos\phi+c_x\sqrt{{\cal D}_0}|}
{2\sqrt{{\cal D}_0}(c_x^2+s_x^2\cos^2\phi)};
{\cal D}_0=c_x^2+s_x^2\cos^2\phi-c_x^{'2}, 
\label{eq:eqs7}
\ee
and
\be
c'_x=\frac{Mx'_{Bj}}{p'R'}\left ( 1- \frac{t_1}{xQ^2}- \frac{t_1}{2M^2x'_{Bj}}
\right ).
\label{eq:eqs7a}
\ee
All kinematical invariants may be obtained from the corresponding ones defined
 above for non shifted kinematics, by replacing $s\to sx$,  $Q^2\to Q^2x$ and  $\varepsilon_-\to \varepsilon_- x$.

\begin{figure}
\includegraphics[scale=.6]{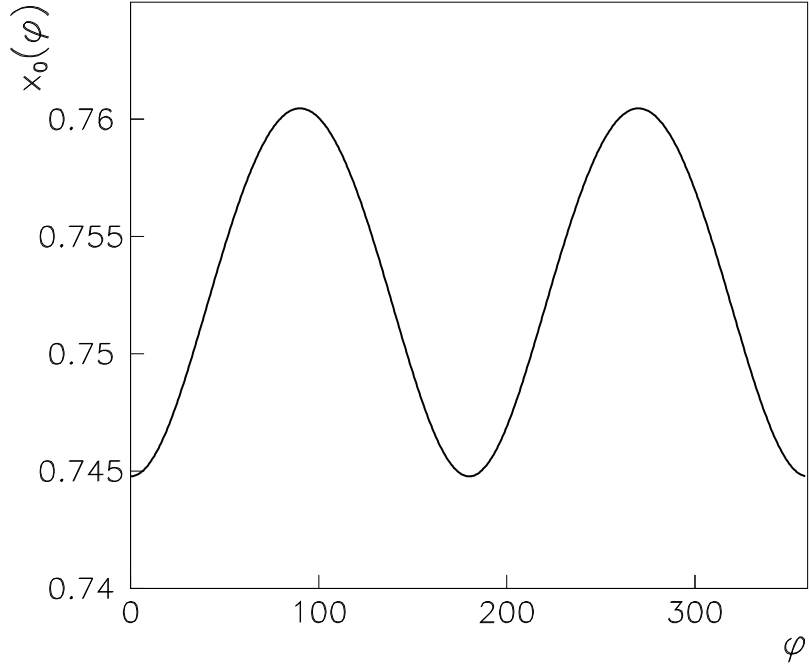}
\caption{ Allowed kinematical region $x>x_0(\phi)$.}
\label{fig:x0}
\end{figure}

The lower limit of integration $x_0(\phi)$ in (\ref{eq:cross:D}) is obtained from the condition:
${\cal D}_0>0$. The curve $x=x_0(\phi)$ is plotted in Fig. \ref{fig:x0}, for the kinematics as in \cite{Mu06}. The kinematically allowed  region is delimited by $x_0(\phi)<x<1$, above the curve.  

\end{document}